\documentclass{cjour}
\usepackage{cjourps}
\newcommand{\lv}[1]{\ensuremath{\mathbf{l}_{#1}}}
\newcommand{\rv}[1]{\ensuremath{\mathbf{r}_{#1}}}
\newcommand{\kv}[1]{\ensuremath{\mathbf{k}_{#1}}}
\newcommand{\dv}[1]{\ensuremath{\mathbf{d}_{#1}}}
\begin{document}









\title{Calculating the $r$$^{-n}$ Interactions, $1\leq n< 3$, in a Periodic
System with a Neutralizing Background Charge Density}


\author{Minghong~G.\ Wu} 

\affil{Chemical Engineering Department,
University of California, Los Angeles, CA 90095-1592, United States}

%







\email{gilwu@sockeye.seas.ucla.edu}


\abstract{The $r^{-n}$ interaction energy, $1\leq n< 3$, for a
infinitely periodic system with explicit charges and a neutralizing,
uniform background charge density is derived. An Ewald based
expression for this energy has an extra term proportional to the
square of the total explicit charges of the system. This expression
may be useful for simulations in which explicit charge neutrality does
not hold or for which the total explicit charges is a fluctuating
quantity.}


\begin{article}


\section{Introduction}
Simulation of atomic systems very often entails the calculation of the
Coulomb interaction of point charges.  The commonly used periodic
boundary conditions embed the simulation box in an infinitely
periodic system. The interaction energy per unit cell volume of this
infinitely periodic system, of $N$ charges $z_i$, $i=1,\ \ldots,\ N$,
can be written as
\begin{eqnarray}
U & = & \frac{1}{2}\sum_{i=1}^N \sum_{j=1}^N 
	{\sum_{\lv{}}}'
	\frac{z_i z_j}{|\rv{i}-\rv{j}-\mathbf{l}|^n}.
\end{eqnarray}
Here $\mathbf{l}$ represents the lattice vectors defined by the
simulation box, and the prime on the
sum excludes the term $\mathbf{l}=\mathbf{0}$ when $i=j$. The Coulomb
interaction corresponds to the case $n=1$. To make the
discussion and derivation slightly more general, the inverse power is
allowed to vary in the range $1\leq n < 3$. It has been shown that when
$1\leq n\leq 3$, the sum converges only when the overall neutrality is
met, i.e., $\sum_{i=1}^N z_i =0$~\cite{Williams_71}. For $n=1$, the
sum is conditionally convergent, i.e., the value of the sum depends on
the order in which the terms are added. An elegant, conventional way
to calculate this sum, the Ewald summation, introduces a
convergence function that separates the sum into two components. The
first component converges rapidly in the real space and the second
converges rapidly in the Fourier
space~\cite{Ewald_21,Nijboer_57,Tosi_64,Frenkel_96}. The
$\kv{}=\mathbf{0}$ term, \kv{} being the reciprocal
lattice vectors,  in
the Fourier space depends on the shape of the infinitely periodic
system~\cite{De_Leeuw_80,Smith_81} and can be recast as a
surface integral~\cite{Deem_90}. Its value can be determined
analytically in simple cases or evaluated numerically in the general
cases. Ignoring this term corresponds to the boundary conditions with
an infinite dielectric constant~\cite{Frenkel_96}. Faster algorithms
for the calculation of the Coulomb interaction have been developed, and
references can be found in a recent review paper~\cite{Schlick_99}.

Problems arise during simulation of charged systems when some ions are
treated implicitly. For example, when a peptide with basic or acidic
amino acids is embedded in a shell of water molecules, the explicit
charges may not meet the neutrality condition and, hence, result in an
infinite energy.  One possible way to overcome this problem is to add
counter ions or fictitious charges in a way to maintain charge
neutrality~\cite{Falconi_98}. This paper presents an alternative way
to arrive at a finite energy, by adding to the system a uniform,
neutralizing charge density $\rho= -\sum_{i=1}^{N}z_i/V_c$, $V_c$
being the volume of the simulation box, or unit cell.  This background
charge density adds negligible amount of computational time, and the
result obtained from this derivation may be useful in practical
simulations. It allows for a computationally convenient approach to
ionic systems where some ions and atoms are treated implicitly. In
particular, it facilitates simulations of ionic systems in the grand
canonical ensemble, where charged molecules can be added to or removed
from the system. The accommodation of the inverse power $n=2$ may
provide useful expressions for screened Coulomb interactions.  For
example, one way to model the solvent effects in the AMBER
forcefield~\cite{Weiner_86,Cornell_95} for biological systems is use a
distance-dependent dielectric constant such as $\epsilon/\epsilon_0 =
4r$, where $r$ is in \AA~\cite{Guenot_92}. Therefore, the screened
Coulomb interaction is proportional to $r^{-2}$ and can be calculated
by the results developed for $n=2$.

This paper is organized as follows. In Sec.~\ref{sec:devel} a Ewald
based, computationally useful expression for the energy of this system
is derived. It is shown that that an extra term proportional to
$(\sum_{i=1}^N z_i)^2$ arises when the explicit charge neutrality does
not hold. The result and the $\kv{}=\mathbf{0}$ term
originating from the boundary conditions 
are discussed in Sec.~\ref{sec:discuss}. The conclusion is given
in Sec.~\ref{sec:conclude}.

\section{Development}
\label{sec:devel}
Consider in a infinitely periodic system the potential $\psi_i$
contributed by a point charge $z_i$ located at \rv{i}, and its
neutralizing uniform density $\rho_i=-z_i/V_c$. For $1\leq n < 3$,
the potential can be written as
\begin{eqnarray}
\label{eqn:poten}
\psi_i(\rv{}) &= & \lim_{L\rightarrow\infty} \left[ z_i \sum_{\lv{}}
	\frac{1}{|\rv{}-\rv{i}-{\lv{}}|^n}\vartheta(L;\ \lv{})
		- \frac{z_i}{V_c}\int
	\frac{1}{|\rv{}-\rv{0}-\rv{}'|^n}\vartheta(L;\
	\rv{}')d\rv{}'\right],\nonumber \\
\end{eqnarray}
where $L$ is the characteristic dimension of the periodic system and
\begin{eqnarray}
\vartheta(L;\ \lv{})&= &\left\{
	\begin{array}{ll}
		1,& \rv{}\in V(L)\\
                0,& \rv{}\notin V(L)
	\end{array}\right.
\end{eqnarray}
is a cutoff function that is added to specify the shape of the
periodic system. The vector
\rv{0} represents the geometrical center of the cell,
$\rv{0} =  \int_{V_c} \rv{} d\rv{}/V_\mathrm{c}$. The procedure of including the shape of the volume
followed by taking the limit is non-trivial and leads to the $\kv{}=\mathbf{0}$
term in the Coulomb interaction case~\cite{Deem_90}. Note that when $n\geq 3$, $\psi_i(\rv{})$
diverges due to the singularity at $\rv{}'= \rv{} -\rv{0}$. 
Next, we introduce a convergence function $\varphi(r)$, which tends
rapidly to zero at infinity.  This convergence function is used to
split $\psi_i(\rv{})$ into two components:
\begin{eqnarray}
\label{eqn:poten_split}
\psi_i^{(R)}(\rv{}) &= & \lim_{L\rightarrow \infty}\left[
		z_i \sum_{\lv{}} \frac{\varphi(|\rv{}-\rv{i}-\lv{}|)}
		           {|\rv{}-\rv{i}-\lv{}|^n}\vartheta(L;\
		\lv{})\right.
\nonumber\\
& &	     -\left.
			\frac{z_i}{V_c}\int
		           \frac{\varphi(|\rv{}-\rv{0}-\rv{}'|)}
				{|\rv{}-\rv{0}-\rv{}'|^n}\vartheta(L;\
		\rv{}')
	d\rv{}'\right]
\nonumber\\
\psi_i^{(F)}(\rv{}) &= &
	    \lim_{L\rightarrow \infty}\left[
		z_i \sum_{\lv{}} \frac{1-\varphi(|\rv{}-\rv{i}-{\lv{}}|)}
		           {|\rv{}-\rv{i}-{\lv{}}|^n}\vartheta(L;\
		\lv{})\right.
\nonumber\\
& &	     -\left.
		\frac{z_i}{V_c}\int
		           \frac{1-\varphi(|\rv{}-\rv{0}-\rv{}'|)}
				{|\rv{}-\rv{0}-\rv{}'|^n}\vartheta(L;\
		\rv{}')d\rv{}'
	\right].\nonumber\\
\end{eqnarray}
Here $\psi_i^{(R)}(\rv{})$ and $\psi_i^{(F)}(\rv{})$ stand for the
short-ranged and long-ranged parts of the sum, respectively. Assuming
the convergence function $\varphi(r)$ decays fast enough so that each
term in $\psi_i^{(R)}(\rv{})$ is finite, the limit
$L\rightarrow \infty$ can be formally taken and
\begin{eqnarray}
\label{eqn:psi_i_r}
\psi_i^{(R)}(\rv{}) &= &
		z_i \sum_{\lv{}} \frac{\varphi(|\rv{}-\rv{i}-\lv{}|)}
		           {|\rv{}-\rv{i}-\lv{}|^n}
	     -
			\frac{z_i}{V_c}\int
		           \frac{\varphi(|\rv{}-\rv{0}-\rv{}'|)}
				{|\rv{}-\rv{0}-\rv{}'|^n}d\rv{}'
\nonumber\\
&= &    		z_i \sum_{\lv{}} \frac{\varphi(|\rv{}-\rv{i}-\lv{}|)}
		           {|\rv{}-\rv{i}-\lv{}|^n}
	     -
			\frac{z_i}{V_c}\int
		           \frac{\varphi(|\rv{}'|)}{|\rv{}'|^n}d\rv{}'
\end{eqnarray}
Note that the average of $\psi_i^{(R)}$ over the simulation box is zero:
\begin{eqnarray}
\label{eqn:psi_i_avg}
\int_{V_c}\psi_i^{(R)}(\rv{})d\rv{} &= & \int_{V_c} z_i \sum_{\lv{}}
	\frac{\varphi(|\rv{}-\rv{i}-\lv{}|)}{|\rv{}-\rv{i}-\lv{}|^n} d\rv{}
		- \int_{V_c}\frac{z_i}{V_c}\int
	\frac{\varphi(|\rv{}'|)}{|\rv{}'|^n}d\rv{}'d\rv{}
\nonumber \\
& =& \int z_i \frac{\varphi(|\rv{}'|)}{|\rv{}'|^n}d\rv{}'
		-
	z_i\int\frac{\varphi(|\rv{}'|)}{|\rv{}'|^n}d\rv{}'
\nonumber \\
& =& 0,
\end{eqnarray}
where $\int_{V_c}$ denotes integration over the simulation box.
The long-ranged $\psi_i^{(F)}(\rv{})$ is best evaluated
by rewriting it as a sum of Fourier components. Let
$F_{\kv{}}\left[f(\rv{})\right]$ denote the Fourier transform
\begin{eqnarray}
F_{\kv{}}\left[f(\rv{})\right]&= &\int f(\rv{})\exp(-i\kv{}\cdot\rv{})d\rv{}.
\end{eqnarray}
The Fourier components of $\psi_i^{(F)}(\rv{})$ are non-zero only at
the reciprocal lattice vectors \kv{}. Therefore,
\begin{eqnarray}
\label{eqn:psi_i_kn0} 
 \psi_i^{(F)}(\rv{}) &= & \frac{z_i}{V_c} \sum_{\kv{}\neq \mathbf{0}}
		F_{\kv{}}\left[\frac{1-\varphi(|\rv{}'|)}{|\rv{}'|^n}\right]
	\exp\left[i\kv{}\cdot(\rv{}-\mathbf{r}_i)\right] +\omega_i(\rv{}),
\end{eqnarray}
where 
\begin{eqnarray}
\label{eqn:psi_i_k0}
\omega_i(\rv{}) &= &
\lim_{L\rightarrow \infty}\frac{z_i}{V_c} \int \left(\frac{1-\varphi(|\rv{}-\rv{i}-\rv{}'|)}
	           {|\rv{}-\rv{i}-\rv{}'|^n}-\frac{1-\varphi(|\rv{}-\rv{0}-\rv{}'|)}
	           {|\rv{}-\rv{0}-\rv{}'|^n}\right)\vartheta(L;\
	           \rv{}')d\rv{}'\nonumber\\
\end{eqnarray}
For $n=1$, $\omega_i(\rv{})$ leads to the non-vanishing
$\kv{}=\mathbf{0}$ term as $L\rightarrow \infty$. However, it diminishes when
$n>1$. We shall neglect this term for now and discuss it later in
Sec.~\ref{sec:discuss}.

Uses of Eqs.~(\ref{eqn:poten_split}), (\ref{eqn:psi_i_r}), and 
(\ref{eqn:psi_i_kn0}) give
\begin{eqnarray}
\label{eqn:psi_i}
 \psi_i(\rv{}) &= &z_i \sum_{\lv{}} \frac{\varphi(|\rv{}-\rv{i}-{\lv{}}|)}
		           {|\rv{}-\rv{i}-{\lv{}}|^n}
		-\frac{z_i}{V_c} \int
		           \frac{\varphi(|\rv{}'|)}{|\rv{}'|^n}d\rv{}' 
\nonumber\\
& &	 + \frac{z_i}{V_c}\sum_{\kv{}\neq \mathbf{0}}
	   F_{\kv{}}\left[\frac{1-\varphi(|\rv{}'|))}{|\rv{}'|^n}\right]
		 \exp\left[i\kv{}\cdot(\rv{}-\mathbf{r}_i)\right].
\end{eqnarray}

Now we add up the contributions from all charges and the background
density and let the total potential be
$\psi(\rv{})=\sum_{i=1}^N\psi_i(\rv{})$.  The
total interaction energy, per unit cell volume, is therefore
\begin{eqnarray}
\label{eqn:total_e}
U & =&   \frac{1}{2}\sum_{i=1}^N z_i
	\left[\psi(\rv{})-\frac{z_i}{|\rv{}-\rv{i}|^n}\right]_{\rv{}=\rv{i}} 
	-\frac{1}{2}\rho\int_{V_c}\psi(\rv{})d\rv{}.
\end{eqnarray}
The bracketed term in
Eq.~(\ref{eqn:total_e}) stands for the potential acting on the point
charge $i$, and can be rewritten as
\begin{eqnarray}
\label{eqn:psi_i_m_self}
\left[\psi(\rv{})-\frac{z_i}{|\rv{}-\rv{i}|^n}\right]_{\rv{}=\rv{i}}
&= & \sum_{j=1}^N{\sum_{\lv{}}}'
     \frac{z_j\varphi(|\rv{i}-\rv{j}-{\lv{}}|)}{|\rv{i}-\rv{j}-{\lv{}}|^n}
\nonumber \\
&&	-z_i\lim_{|\rv{}| \rightarrow 0} 
			\frac{1-\varphi(|\rv{}|)}{|\rv{}|^n}
	-\frac{\sum_{j=1}^N z_j}{V_c}\int\frac{\varphi(|\rv{}'|)}{|\rv{}'|^n}d\rv{}'
\nonumber \\
&&\!\!	+\frac{1}{V_c}\sum_{\kv{}\neq \mathbf{0}}
	 F_{\kv{}}\left[\frac{1-\varphi(|\rv{}|)}{|\rv{}|^n}\right]
		 \exp\left(i\kv{}\cdot\rv{i}\right)S(\kv{})
\end{eqnarray}
Here $S(\mathbf{k})=\sum_{j=1}^N z_j\exp(-i\mathbf{k}\cdot\rv{j})$ is
the structure factor. If we ignore the surface contribution, the
$\kv{}=\mathbf{0}$ term is zero, as is the second term at the right
hand side of Eq.~(\ref{eqn:total_e}).  Finally, substituting
Eq.~(\ref{eqn:psi_i_m_self}) into Eq.~(\ref{eqn:total_e}) yields an
expression for the total energy:
\begin{eqnarray}
\label{eqn:total_e_gen}
U  & =&\frac{1}{2}\sum_{i=1}^N \sum_{j=1}^N {\sum_{\lv{}}}'
	\frac{z_i z_j\varphi(|\rv{i}-\rv{j}-{\lv{}}|)}{|\rv{i}-\rv{j}-{\lv{}}|^n}
	-\frac{(\sum_{i=1}^N z_i)^2}{2V_c}
	 	\int\frac{\varphi(|\rv{}'|)}{|\rv{}'|^n}d\rv{}'
\nonumber \\
  &  &	-\frac{\sum_{i=1}^N {z_i}^2}{2}
		\lim_{|\rv{}| \rightarrow 0} 
			\frac{1-\varphi(|\rv{}|)}{|\rv{}|^n}
        +\frac{1}{2V_c}\sum_{\kv{}\neq \mathbf{0}}
		\left|S(\kv{})\right|^2
	 	F_{\kv{}}\left[\frac{1-\varphi(|\rv{}|)}{|\rv{}|^n}\right]
\end{eqnarray}

The convergence function $\varphi(r)$ should be chosen in a way that
each term of Eq.~(\ref{eqn:total_e_gen}) is finite and computable. One
possible choice is $\varphi(r)=\frac{\Gamma(n/2,\ \alpha
r^2)}{\Gamma(n/2)}$~\cite{Williams_71}, where $\Gamma(n/2)$ and
$\Gamma(n/2,\ \alpha r^2)$ are the gamma function and the incomplete
gamma function, respectively. The parameter $\alpha$ adjusts the
contributions from the real and the reciprocal parts and should be
optimized for the system of interest. With this choice of
$\varphi(r)$, the Fourier transform of
$(1-\varphi(|\rv{}|))/|\rv{}|^n$ has been shown to be~\cite{Nijboer_57}
\begin{eqnarray}
F_{\kv{}}\left[\frac{1-\varphi(|\rv{}|)}{|\rv{}|^n}\right]
 &=&
	\frac{2^{3-n}\pi^{3/2}}{k^{3-n}\Gamma(n/2)}
	\Gamma\left(\frac{3-n}{2},\ \frac{k^2}{4\alpha}\right)
\end{eqnarray}
and
\begin{eqnarray}
\lim_{|\rv{}| \rightarrow 0} 
			\frac{1-\varphi(|\rv{}|)}{|\rv{}|^n}
&= & \frac{2\alpha^{n/2}}{n\Gamma(n/2)}.
\end{eqnarray}
The short-ranged part of the contribution from the background charge
density can be evaluated analytically to be
\begin{eqnarray}
\int\frac{\varphi(|\rv{}|)}{|\rv{}|^n}d\rv{} & =&
	\frac{2\pi^{3/2}}{(3-n)\Gamma(n/2)\alpha^{(3-n)/2}},\ 0<n<3.
\end{eqnarray}
Therefore, the interaction energy can be written as
\begin{eqnarray}
\label{eqn:U_useful}
U  & =&
\frac{1}{2\Gamma(n/2)}\left[
	\sum_{i=1}^N \sum_{j=1}^N {\sum_{\lv{}}}'
	\frac{z_i z_j\Gamma(\frac{n}{2},\ \alpha|\rv{i}-\rv{j}-{\lv{}}|^2)}
	{|\rv{i}-\rv{j}-{\lv{}}|^n}\right.\nonumber\\
& &
	-\frac{2\pi^{3/2}}{(3-n)V_c\alpha^{(3-n)/2}}\left(\sum_{i=1}^N z_i\right)^2
-\frac{2\alpha^{n/2}}{n}\sum_{i=1}^N {z_i}^2\nonumber\\
& &
	+\left.\frac{2^{3-n}\pi^{3/2}}{V_c}\sum_{\kv{}\neq \mathbf{0}}
		\frac{|S(\mathbf{k})|^2}{k^{3-n}}	
			\Gamma\left(\frac{3-n}{2},\ \frac{k^2}{4\alpha}\right)\right].
\end{eqnarray}
In the Coulomb interaction case, $n=1$, $\varphi(r)$ is the complementary
error function $\mathrm{erfc}(\alpha^{1/2}r)$ and
Eq.~(\ref{eqn:U_useful}) is reduced to the Ewald
sum~\cite{Frenkel_96} plus an extra term proportional to
$(\sum_{i=1}^N z_i)^2$:
\begin{eqnarray}
U_{n=1}  & =&
\frac{1}{2}\sum_{i=1}^N \sum_{j=1}^N {\sum_{\lv{}}}'
z_i z_j
\frac{\mathrm{erfc}\left(\alpha^{1/2}|\rv{i}-\rv{j}-\lv{}|\right)}{|\rv{i}-\rv{j}-\lv{}|}
-\frac{\pi}{2V_c\alpha}\left(\sum_{i=1}^N z_i\right)^2\nonumber\\
&&
-\left(\frac{\alpha}{\pi}\right)^{1/2}\sum_{i=1}^N {z_i}^2
+\frac{2\pi}{V_c}\sum_{\kv{}\neq \mathbf{0}}
		\frac{|S(\mathbf{k})|^2}{k^2}\exp\left(-\frac{k^2}{4\alpha}\right).
\end{eqnarray}
For $n=2$, $\psi(r) = \exp(-\alpha r^2)$ and the corresponding equation is
\begin{eqnarray}
U_{n=2}  & =&
\frac{1}{2}\sum_{i=1}^N \sum_{j=1}^N {\sum_{\lv{}}}'
z_i z_j
\frac{\exp(-\alpha |\rv{i}-\rv{j}-\lv{}|^2)}{|\rv{i}-\rv{j}-\lv{}|^2}
-\frac{\pi^{3/2}}{V_c\alpha^{1/2}}\left(\sum_{i=1}^N z_i\right)^2\nonumber\\
&&
- \frac{\alpha}{2}\sum_{i=1}^N {z_i}^2
+\frac{\pi^2}{V_c}\sum_{\kv{}\neq \mathbf{0}}
		\frac{|S(\mathbf{k})|^2}{k}\mathrm{erfc}\left(\frac{k}{2\alpha^{1/2}}\right).
\end{eqnarray}

\section{Discussion}
\label{sec:discuss}
The $\kv{}=\mathbf{0}$ term described by Eq.~(\ref{eqn:psi_i_k0}) does
not diminish when $n\leq 1$. Let $\omega(\rv{})=\sum_{i=1}^N
\omega_i(\rv{})$. When $\omega(\rv{})$ is multipole-expanded and the
limit $L\rightarrow \infty$ is taken, the
first term becomes zero due to charge neutrality. Thus,
the leading term in $\omega(\rv{})$ is the second term, which
is non-zero only at the surface of the periodic system and can be recast as
\begin{eqnarray}
\label{eqn:surface_psi}
\omega(\rv{}) &= & \lim_{L\rightarrow\infty}\frac{n}{V_c}\int_{S(L)} 
	\frac{(\rv{}-\rv{\mathrm{M}})\cdot\rv{}'\, \dv{}\cdot\hat{\mathbf n}}{{\rv{}'}^{n+2}}\,dS,
\end{eqnarray}
where 
\begin{eqnarray}
\rv{\mathrm{M}} &= &\frac{\sum_{i=1}^N|z_i|\rv{i} + |\sum_{i=1}^N z_i|\rv{0}}
		   {\sum_{i=1}^N|z_i|+ |\sum_{i=1}^N z_i|}
\end{eqnarray}
is the ``center of mass'' of the simulation box, $\dv{} = \sum_{i=1}^N
z_i (\rv{i} - \rv{0})$ is the dipole moment of the simulation box, and
$S(L)$ is the surface of the volume with a characteristic dimension
$L$. Note that both the explicit charges and the background charge
density contribute to the dipole moment. It is interesting to see
that $\omega(\rv{})$ is the only term that contributed to the average
potential over the simulation box, i.e.,
\begin{eqnarray}
\label{eqn:average_psi}
\frac{1}{V_c}\int_{V_c}\psi(\rv{}) &= &
\frac{1}{V_c}\int_{V_c}\omega(\rv{}) \nonumber\\
&= &\lim_{L\rightarrow\infty}\frac{n}{V_c}\int_{S(L)} 
	\frac{(\rv{0}-\rv{\mathrm{M}})\cdot\rv{}'\, \dv{}\cdot\hat{\mathbf n}}{{\rv{}'}^{n+2}}\, dS.
\end{eqnarray}
Note that the average of both the short-ranged part and the $\kv{}\neq 0$ terms over
the simulation box is zero.

The contribution of $\omega(\rv{})$ to the energy per unit
cell is
\begin{eqnarray}
\label{eqn:surface_U}
  U_\mathrm{s}
  &=  &  \frac{1}{2}\sum_{i=1}^N z_i\omega(\rv{i})-\frac{1}{2}\rho\int_{V_c}\omega(\rv{})d\rv{}.
\end{eqnarray}
Substituting the expression for $\omega(\rv{})$ into
Eq.~(\ref{eqn:surface_U}) gives
\begin{eqnarray}
\label{eqn:surface_int}
U_\mathrm{s} &= & \lim_{L\rightarrow\infty}\frac{n}{2V_c}\int_{S(L)}
\frac{\dv{}\cdot\rv{}'\, \dv{}\cdot\hat{\mathbf n}}{{\rv{}'}^{n+2}}\, dS.
\end{eqnarray}
Clearly, this term diverges when $n<1$ and diminishes
when $n>1$ as $L\rightarrow \infty$. For $n=1$ it contributes a finite
energy to the system~\cite{Deem_90}. It is interpreted as the
interaction between the dipole moment of the system and the surface
charges at the boundary. Ignoring this term corresponds to using
boundary conditions with an infinite dielectric constant.

It is worth mentioning that, for the $n>3$ case, an expression having
the same form as Eq.~(\ref{eqn:U_useful}) was derived in
ref.~\cite{Williams_71} without the addition of a neutralizing
background charge density. Here the same equation is extended to the
case $1\leq n < 3$ with a different physical interpretation.

\section{Conclusion}
\label{sec:conclude}
The $r^{-n}$ interaction energy, $1\leq n< 3$, for a infinitely
periodic system with charges and a neutralizing, uniform background
charge density is derived. Following the Ewald transformation, the
interaction energy is split into a real space sum and Fourier space
sum, and both are rapidly convergent.  The Ewald based expression for
this energy has an extra term proportional to the square of the total
explicit charge of the system. This expression may be useful for
simulations in which explicit charge neutrality does not hold or for
which the total explicit charges is a fluctuating quantity.







\begin{acknowledgment}
This research was supported by the National Science Foundation.
\end{acknowledgment}


\end{article}
\end{document}